\documentclass{article}
\usepackage{arxiv}
\usepackage[utf8]{inputenc} 
\usepackage[T1]{fontenc}    
\usepackage{hyperref}       
\usepackage{url}            
\usepackage{booktabs}       
\usepackage{amsfonts}       
\usepackage{nicefrac}       
\usepackage{microtype}      
\usepackage{lipsum}
\usepackage{float}
\usepackage{amsmath}
\usepackage{amsmath}
\usepackage{threeparttable}
\usepackage{graphicx}
\usepackage{geometry} 
\usepackage{amsmath} 
\usepackage{amssymb} 

\begin{document}

\title{Generalized Autoregressive Score asymmetric Laplace Distribution and Extreme Downward Risk Prediction}

\author{
  Hong Shaopeng\footnote{e-mail: hsp\_cueb@163.com} \\
  School of Statistics\\
  Capital University of Economics and Trade\\
  BeiJing, China
}

\maketitle

\begin{abstract}
Due to the skessed distribution, high peak and thick tail and asymmetry of financial return data, it is difficult to describe the traditional distribution. In recent years, generalized autoregressive score (GAS) has been used in many fields and achieved good results. In this paper, under the framework of generalized autoregressive score (GAS), the asymmetric Laplace distribution (ALD) is improved, and the GAS-ALD model is proposed, which has the characteristics of time-varying parameters, can describe the peak thick tail, biased and asymmetric distribution. The model is used to study the Shanghai index, Shenzhen index and SME board index. It is found that: 1) the distribution parameters and moments of the three indexes have obvious time-varying characteristics and aggregation characteristics. 2) Compared with the commonly used models for calculating VaR and ES, the GAS-ALD model has a high prediction effect.
\end{abstract}


\section{Introduction}
As a special kind of economic system, the financial market has emerged with a series of unique macro laws, which are usually called "stylized facts". For example, the return distribution is sharp and asymmetrical. Due to Knight uncertainty, static parameter models usually cannot accurately describe the characteristics of financial returns, and dynamic parameter models are needed to characterize the distribution of returns.

The structure of this paper is as follows: The first part reviews the research results of asymmetric Laplacian distribution and generalized autoregression at home and abroad, the second part builds the Laplacian distribution based on the generalized autoregressive score, and the third part uses generalized autoregressive The Plass distribution conducts empirical research on China's Shanghai Stock Index, Shenzhen Index and SME Board Index. The fourth part is conclusions and recommendations.
\section{literature}
\subsection{Discussion on the distribution of returns}
In risk management, VaR and ES have become more reliable risk measurement indicators, and how to calculate VaR and ES more effectively has always been a hot issue in risk management. The core of calculating VaR and ES is to describe the distribution of its rate of return. However, as a special kind of economic system, the financial system has many unique and typical facts. For example, financial returns have typical facts such as volatility clustering, autocorrelation, spikes and fat tails. It is unreasonable to assume that the rate of return is normally distributed.  Bollerslev (1987) \cite{bib1} proposed to use the t distribution to characterize the fat tail characteristics of the return distribution. Often financial market returns show right-skewed characteristics, and normal distribution and t-distribution, as distributions with zero skewness, cannot be described. Hansen (1994)\cite{bib2} proposed to use the biased t distribution to simultaneously characterize the fat tail and biased characteristics of the return distribution. However, in the financial market, the asymmetry of gains and losses makes the distribution of returns appear asymmetry in the left and right tails. Zhu Dongming and Galbraith (2010)\cite{bib3}  proposed two asymmetric t distributions with higher degrees of freedom than the biased t distribution based on the biased t distribution. These two t distributions can better characterize the asymmetry of the return rate by controlling the asymmetry of the left and right tails. Keming Yu and Zhang (2005) \cite{bib4} proposed an asymmetric Laplace distribution (ALD) controlled by three parameters. Compared with other complex distributions (such as generalized error distribution), ALD has fewer parameters, and each order moment has a clear display expression.

Due to the superiority of ALD, a large number of scholars have conducted research on ALD. Liu Jianyuan and Liu Qiongsun (2007)\cite{bib5} used ALD to fit the returns of China’s stock market, and compared the symmetric Laplace distribution, the normal distribution and the asymmetric Laplace distribution to calculate the effectiveness of VaR. The Plass distribution has the best fitting result. Du Hongjun and Wang Zongjun (2013) \cite{bib6} used the asymmetric Laplacian distribution to calculate the dynamic risk VaR and CVaR. The asymmetric Laplacian distribution can well describe the characteristics of bias, spikes and fat tails. Liu Pan and Zhou Ruomei (2015) \cite{bib7} used AEPD, AST and ALD distributions to describe the typical facts of financial asset returns, and found that the ALD distribution is significantly better than other distributions in calculating the long VaR value of the bottom quantile. . Based on ALD, Tang Jianyang et al. (2018) \cite{bib8} proposed the AR-GJR-GARCH model with ALD error distribution, which has a better performance in calculating VaR and ES than traditional models that measure VaR and ES. 
\subsection{The Proposal and Application of Generalized Autoregressive Score}
Creal (2012)\cite{bib9} proposed the Generalized Autoregressive Scoring Model (GAS) as a framework for time-varying parameter modeling, which has been widely used in recent years. Such as GAS-EWMA (Lucas, 2016\cite{bib10}), beta-t-EGARCH (Harvey and Sucarrat, 2014\cite{bib11}), GAS-Copula (Oh and Patton, 2016\cite{bib12}), GAS-factor Copula (leaf five First Class, 2018\cite{bib13}), GAS-D-Vine-Copula (Zou Yumei, 2018\cite{bib14}), GJR-GARCH-skew t-GAS-copula (Xiao-Li Gong, 2019\cite{bib15}) and GAS-EGARCH -EGB2 (Yao Ping et al., 2019\cite{bib16}). The above models have performed well in the field of risk management.

It can be seen from the literature review that the asymmetric Laplacian distribution has a good performance in fitting the rate of return, which can characterize the typical facts of the peak, thick tail, bias and asymmetry of the rate of return, and are similar. Compared with the normal distribution and the t distribution, the asymmetric Laplacian distribution has a better performance in calculating VaR and ES. However, it is impossible to describe the time-varying nature of each order moment of the rate of return, and there are defects in the empirical study. In recent years, the GAS model has been proposed as a time-varying parameter modeling framework. A large number of documents have proved that using the GAS framework to make the model dynamic can improve the performance of the model. In summary, this article considers expanding the asymmetric Laplace distribution under the GAS framework to make its parameters dynamic, and strive to better describe the distribution of returns and more accurate predictions of extreme risks.

The innovations made in this paper are as follows: 1) Use the GAS model to extend the ALD distribution to the GAS-ALD distribution. 2) Apply GAS-ALD to my country's securities market to dynamically measure conditional mean, conditional volatility, conditional skewness and conditional kurtosis. 3) Compare the GAS-ALD model with mainstream VaR and ES models. On the one hand, it enriches the calculation methods of VaR and ES, on the other hand, it can provide investors with investment basis. It has theoretical and practical significance.
\section{ Models}
\subsection{Generalized Autoregressive Score}
The generalized autoregressive scoremodel is a data-driven model. It is a unified framework for time-varying parameter modeling. Compared with other models, GAS has the advantage that it makes full use of distributed information to construct a time-varying parameter evolution process.

Suppose the conditional density of the rate of return y at time t is:

\begin{equation}
\label{1}
\mathbf{y}_t|\mathbf{y}_{1:t-1}\sim f\left(\mathbf{y}_t;\mathbf{\theta}_t\right)\ \ \ \ 
\end{equation}
Among them, $\mathbf{y}_{1:t-1}=(y_1,\ldots,y_{t-1})$, which means all the information of y in the past moment, $\mathbf{\theta}_t $ Is a time-varying parameter set. From the formula (\ref{1}), it can be seen that the conditional distribution only depends on the information at the past moment and the parameter set. The core of the GAS model is the parameter evolution process of time-varying parameters ((\ref{2}) ).

\begin{equation}
\label{2}
\mathbf{\theta}_{t+1}\equiv\mathbf{\kappa}+\mathbf{A}\mathbf{s}_t+\mathbf{B}\mathbf{\theta}_t
\end{equation}

Among them, $\mathbf{\kappa}$ is a constant matrix, A and B are coefficient matrices, then the score $s_t$ is the driving mechanism, usually defined as (\ref{3}) and (\ref{4}) , (\ref{5}) and (\ref{6}). $\mathcal{I}_t\left(\mathbf{\theta}_t\right)$ is the Fisher information matrix of the time-varying parameter set. $\gamma$ is usually in $\{0,\frac{1}{2 },1\}$, used to adjust the score to calculate the final update item $s_t$, this article selects $\gamma=0$, that is, $s_t=\mathrm{\nabla}_t$. It can be seen that the parameter evolution in the GAS model is divided into two parts, the first part is driven by the score function, and the second part is the AR(1) process.

\begin{equation}
\label{3}
s_t=\mathbf{S}_t\left(\mathbf{\theta}_t\right)\mathrm{\nabla}_t\left(\mathbf{y}_t,\mathbf{\theta}_t\right)
\end{equation}
\begin{equation}
\label{4}
\nabla_{t}\left(\mathbf{y}_{t}, \theta_{t}\right) \equiv \frac{f\left(\mathbf{y}_{t} ; \theta_{t}\right)}{\partial \theta_{t}}
\end{equation}
\begin{equation}
\label{5}
\mathbf{S}_t\left(\mathbf{\theta}_t\right)\equiv\mathcal{I}_t\left(\mathbf{\theta}_t\right)^{-\gamma}
\end{equation}

\begin{equation}
\label{6}
\mathcal{J}_{t}\left(\boldsymbol{\theta}_{t}\right) \equiv \mathrm{E}_{t-1}\left[\boldsymbol{\nabla}_{t}\left(\mathbf{y}_{t}, \boldsymbol{\theta}_{t}\right) \boldsymbol{\nabla}_{t}\left(\mathbf{y}_{t}, \boldsymbol{\theta}_{t}\right)^{\top}\right]
\end{equation}
\subsection{Asymmetric Laplace distribution}
The probability density function of the asymmetric Laplace distribution is as (\ref{den}):
\begin{equation}
\label{den}
f(x ; \mu, \sigma, p)=\frac{p}{\sigma(1+p^{2})}\left\{\begin{array}{ll}
\exp \left(-\frac{\mathrm{p}}{\sigma}(x-\mu)\right), &  \mathrm{x} \geq \mu \\
\exp \left(\frac{1}{\sigma \mathrm{p}}(x-\mu)\right), &  \mathrm{x}<\mu
\end{array}\right.
\end{equation}
The expressions of the moments of ALD are as follows:
\begin{equation}
E(X)=\mu+\frac{\sigma}{\sqrt2}(\frac{1}{p}-p)\ 
\end{equation}
\begin{equation}
\operatorname{Var}(\mathrm{X})=\left[\frac{\sigma}{\sqrt{2}}\left(\frac{1}{\mathrm{p}}-\mathrm{p}\right)\right]+\sigma^{2}
\end{equation}
\begin{equation}
\text { Skewness }=2 \frac{1 / p^{3}-p^{3}}{\left(1 / p^{2}+p^{2}\right)^{3 / 2}}
\end{equation}
\begin{equation}
\text { Kurtosis }=6-\frac{12}{\left(1 / p^{2}+p^{2}\right)^{2}}
\end{equation}

\begin{figure}
	\centering
	\includegraphics[width=0.6\linewidth]{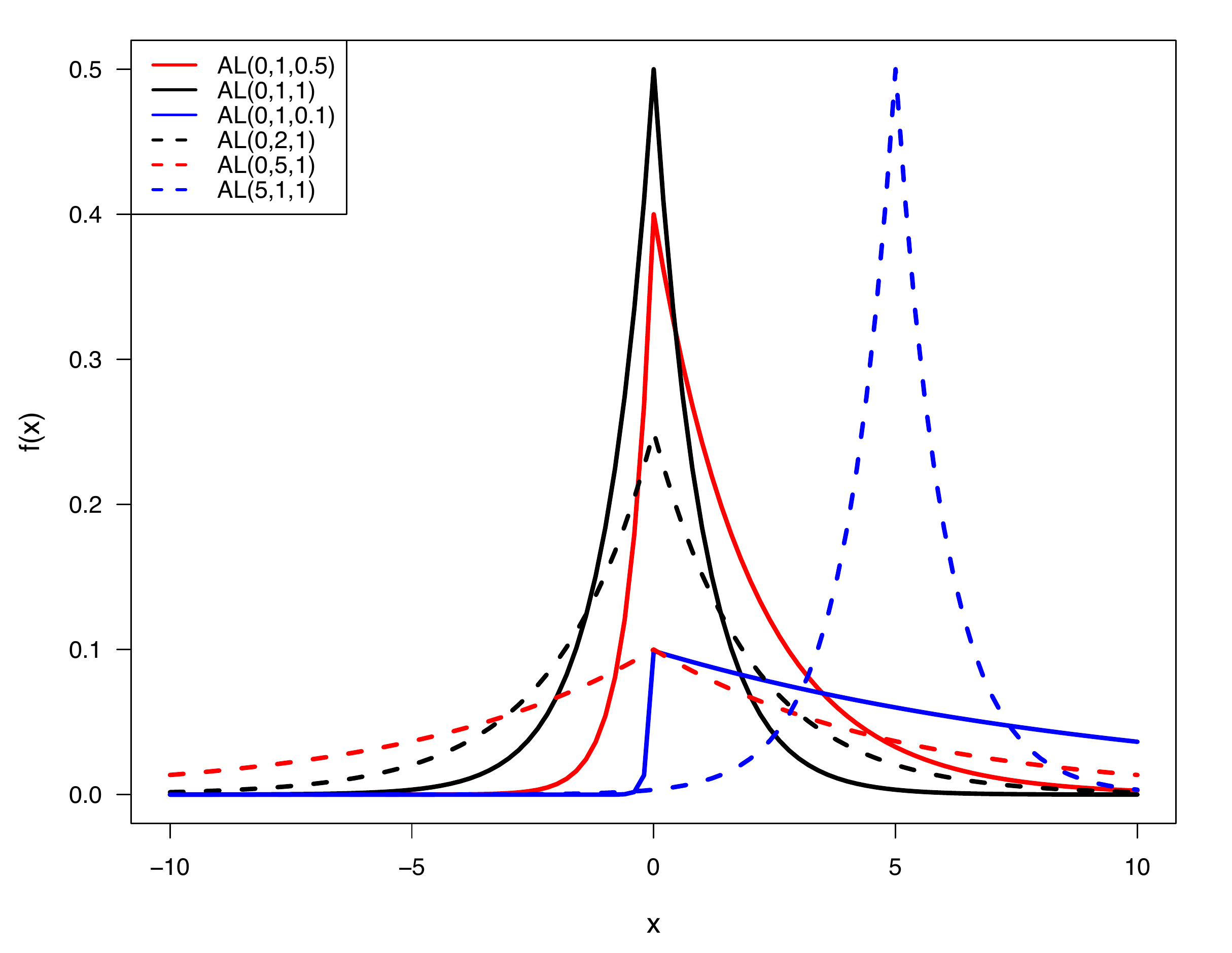}
	\caption{ALD probability density}
	\label{fig:al}
\end{figure}
\subsection{ Generalized Autoregressive Score asymmetric Laplace Distribution}
From the (\ref{1}), rewrite the (\ref{den}) to make its parameters $\mu$, $\sigma$ and $p$ time-varying:
\begin{equation}
\label{10}
\mathbf{y}_{t} \mid \mathbf{y}_{1: t-1} \sim f\left(\mathbf{y}_{t}; \mu_{t}, \sigma_{t }, p_{t}\right)
\end{equation}
\begin{equation}
\label{11}
f(y_t; \mu_t, \sigma_t, p_t)=\frac{p_t}{\sigma_t(1+p_t^{2})}\left\{\begin{array}{ll}
\exp \left(-\frac{\mathrm{p_t}}{\sigma}(y_t-\mu_t)\right), &  \mathrm{x_t} \geq \mu_t \\
\exp \left(\frac{1}{\sigma_t \mathrm{p_t}}(y_t-\mu_t)\right), &  \mathrm{y_t}<\mu_t
\end{array}\right.
\end{equation}
The likelihood function of the asymmetric Laplace distribution is as follows:
\begin{equation}
\ln f=\ln p_{\mathrm{t}}+\ln \left(1+\mathrm{p}_{\mathrm{t}}^{2}\right)-\ln \left(\sigma_ {\mathrm{t}}\right)+\left\{\begin{array}{ll}
-\frac{\mathrm{p}_{\mathrm{t}}}{\sigma}\left(\mathbf{y}_{\mathrm{t}}-\mu_{\mathrm{t}}\right ), &  \mathbf{y}_{\mathrm{t}} \geq \mu \\
\frac{1}{\sigma \mathrm{p}_{\mathrm{t}}}\left(\mathbf{y}_{\mathrm{t}}-\mu_{\mathrm{t}}\right ), &  \mathbf{y}_{\mathrm{t}}<\mu_{\mathrm{t}}
\end{array}\right.
\end{equation}
Set the parameter set to $\theta_t=\{\mu_t,\sigma_t,p_t\}$. Calculate the score function:
\begin{equation}
\nabla_{t}\left(\boldsymbol{y}_{t}, \mu_{t}\right)=\frac{\partial f\left(\boldsymbol{y}_{t}; \mu_{t }\right)}{\partial \mu_{t}}=\left\{\begin{array}{ll}
\frac{\mathrm{p}_{\mathrm{t}}}{\sigma_{\mathrm{t}}}, &  \mathbf{y}_{\mathrm{t}} \geq \mu_{\mathrm{t}} \\
-\frac{1}{\sigma_{\mathrm{t}} \mathrm{p}_{\mathrm{t}}}, &  \mathbf{y}_{\mathrm{t}} <\mu_{\mathrm{t}}
\end{array}\right.
\end{equation}
\begin{equation}
\nabla_{t}\left(\mathbf{y}_{t}, \boldsymbol{\sigma}_{t}\right)=\frac{\partial \log f\left(\mathbf{y}_{ t}; \boldsymbol{\sigma}_{t}\right)}{\partial \boldsymbol{\sigma}_{t}}=-\frac{1}{\boldsymbol{\sigma}_{t}} +\left\{\begin{array}{ll}
\mathrm{p}_{t}\left(\mathrm{y}_{\mathrm{t}}-\mu\right) \boldsymbol{\sigma}_{\mathrm{t}}^{-2} , &  \mathbf{y}_{\mathrm{t}} \geq \mu_{\mathrm{t}} \\
-\frac{1}{\mathrm{p}_{\mathrm{t}}}\left(\mathbf{y}_{\mathrm{t}}-\mu\right) \boldsymbol{\sigma}_ {\mathrm{t}}^{-2}, &  \mathbf{y}_{\mathrm{t}}<\mu_{\mathrm{t}}
\end{array}\right.
\end{equation}
\begin{equation}
\nabla_{t}\left(\mathbf{y}_{t}, p_{\mathrm{t}}\right)=\frac{\partial \log f\left(\mathbf{y}_{t} ; p_{\mathrm{t}}\right)}{\partial \kappa_{\mathrm{t}}}=\frac{1}{p_{\mathrm{t}}}+\frac{2 p_{\ mathrm{t}}}{1+p_{\mathrm{t}}^{2}}+\left\{\begin{array}{ll}
-\frac{1}{\sigma_{\mathrm{t}}}\left(\mathbf{y}_{t}-\mu\right), &  \mathbf{y}_{t } \geq \mu_{t} \\
-\frac{1}{\sigma_{\mathrm{t}}}\left(\mathbf{y}_{t}-\mu\right) p_{\mathrm{t}}^{-2}, &  \mathbf{y}_{t}<\mu_{t}
\end{array}\right.
\end{equation}
\begin{equation}
\nabla_{t}=\left[\nabla_{t}\left(\mathbf{y}_{t}, \mu_{t}\right), \nabla_{t}\left(\mathbf{y}_ {t}, \boldsymbol{\sigma}_{t}\right), \nabla_{t}\left(\mathbf{y}_{t}, \mathbf{p}_{t}\right)\right ]
\end{equation}
The formula (\ref{2}) can be written as:
\begin{equation}
\label{17}
\left(\begin{array}{c}
\mu_{t+1} \\
\sigma_{t+1} \\
\mathbf{p}_{t+1}
\end{array}\right)=\left(\begin{array}{c}
\mathbf{\kappa}_{1} \\
\mathbf{\kappa}_{2} \\
\mathbf{\kappa}_{3}
\end{array}\right)+\left(\begin{array}{ccc}
a_{1} & 0 & 0 \\
0 & a_{2} & 0 \\
0 & 0 & a_{3}
\end{array}\right)\left(\begin{array}{c}
\nabla_{t}\left(\mathbf{y}_{t}, \mu_{t}\right) \\
\nabla_{t}\left(\mathbf{y}_{t}, \boldsymbol{\sigma}_{t}\right) \\
\nabla_{t}\left(\mathbf{y}_{t}, \mathbf{p}_{t}\right)
\end{array}\right)+\left(\begin{array}{ccc}
\mathbf{b}_{1} & 0 & 0 \\
0 & \mathrm{b}_{2} & 0 \\
0 & 0 & \mathrm{b}_{3}
\end{array}\right)\left(\begin{array}{c}
\mu_{t} \\
\boldsymbol{\sigma}_{\mathrm{t}} \\
\mathbf{p}_{t}
\end{array}\right)
\end{equation}
(\ref{10}), (\ref{11}) and (\ref{17}) are the GAS-ALD models.
\subsection{Parameter Estimation}
Compared with other models, the model under the GAS framework can accurately estimate the parameter $\theta_t$ after giving the past information and parameter vector $\phi={\mathbf{\kappa},A,B}$.
Record $y_{1:T}$ as the sample $T$ in $y_t$, and $\phi$ can be  estimate by MLE(\ref{18}) 

\begin{equation}
\label{18}
\begin{aligned}
&\widehat{\phi} \equiv \underset{\phi}{\operatorname{argmax}} \mathcal{L}\left(\phi ; y_{1: T}\right)\\
&\text {S.T. } \mathcal{L}\left(\phi ; y_{1: T}\right)=\log f\left(y_{1} ; \theta_{1}\right)+\sum_{t=2}^{T} \log f\left(y_{t} ; \theta_{T}\right)
\end{aligned}
\end{equation}
When t=1, $\theta_1=(I-B)^{-1}\mathbf{\kappa}$.

The following uses the BFGS algorithm to solve the likelihood function to estimate $\widehat{\phi}$.
\subsection{Risk prediction-VaR and ES}
When making a step forward prediction, we can use the GAS-ALD model to estimate the distribution of the return rate $\mathbf{y}_{t+1}$, thereby predicting VaR and ES one step forward.

\begin{equation}
\mathbf{y}_{t+1} \mid \mathbf{y}_{1: t} \sim f\left(\mathbf{y}_{t+1} ; \mu_{t+1}, \sigma_{t+1}, p_{t+1}\right)
\end{equation}
\begin{equation}
\left(\begin{array}{c}
\mu_{t+1} \\
\sigma_{t+1} \\
\mathbf{p}_{t+1}
\end{array}\right)=\left(\begin{array}{c}
\mathbf{\kappa}_{1} \\
\mathbf{\kappa}_{2} \\
\mathbf{\kappa}_{3}
\end{array}\right)+\left(\begin{array}{ccc}
a_{1} & 0 & 0 \\
0 & a_{2} & 0 \\
0 & 0 & a_{3}
\end{array}\right)\left(\begin{array}{c}
\nabla_{t}\left(\mathbf{y}_{t}, \mu_{t}\right) \\
\nabla_{t}\left(\mathbf{y}_{t}, \boldsymbol{\sigma}_{t}\right) \\
\nabla_{t}\left(\mathbf{y}_{t}, \mathbf{p}_{t}\right)
\end{array}\right)+\left(\begin{array}{ccc}
\mathrm{b}_{1} & 0 & 0 \\
0 & \mathrm{b}_{2} & 0 \\
0 & 0 & \mathrm{b}_{3}
\end{array}\right)\left(\begin{array}{c}
\mu_{t} \\
\boldsymbol{\sigma}_{\mathrm{t}} \\
\mathbf{p}_{t}
\end{array}\right)
\end{equation}

The essence of VaR is a quantile, which can be easily obtained by using the inverse function of the distribution function. When using the GAS-ALD model, since the parameters are time-varying, the dynamic VaR can be obtained.

\begin{equation}
V a R_{t+1}(\alpha) \equiv F^{-1}\left(\alpha ; \theta_{t+1}\right)
\end{equation}

Because VaR does not satisfy the subadditivity, we use ES to supplement VaR. ES, as the average part exceeding VaR, can be obtained using the formula (\ref{es}). Same as VaR, the result of ES is also dynamic.

\begin{equation}
\label{es}
E S_{t+1}(\alpha) \equiv \frac{1}{\alpha} \int_{-\infty}^{V a R_{t+1}} z d F\left(z, \theta_{t+1}\right)
\end{equation}

\subsection{Backtesting}
For the statistical test of VaR, this article mainly uses UC test\cite{bib18} , CC test\cite{bib19}  and DQ test\cite{bib20} . The above methods are relatively mature, so I will not repeat them in this article. Because ES is generally too large relative to VaR, the method of testing the backtest effect of VaR cannot be used to test the backtest effect of ES. In this paper, the bootstrap test proposed in the Mcneil et al.(2000) \cite{bib21} is used to evaluate ES efficacy. The main idea of this test is to construct multiple bootstrap residual samples and use them to calculate test statistics and p-values.

\subsubsection{Bootstrap}
The main steps of the test are as follows\cite{bib22,bib23}:

First define the excess residual $e_t$:

\begin{equation}
\ e_t=\frac{x_t-ES_t^q}{\sigma_t^{1/2}}
\end{equation}
Among them, $x_t$ is the true rate of return exceeding VaR, and $\sigma_t$ is the sum of variance of $x_t-ES_t^q$.

Then define the initial residual sequence $l_t$:
\begin{equation}
{\ \ \ \ \ \ \ l}_t=e_t-\frac{1}{z}\sum_{\eta=1}^{z}e_\eta\
\end{equation}
Among them, $z$ is the number of samples exceeding the residual $e_t$.
Assuming that the mean of $l_t$ is $\overline{l}$ and the standard deviation is $\sigma_l$, define the test statistic: $\delta(l)$.
\begin{equation}
\ \ \delta(l)=\frac{\overline{l}}{\sigma_l}\ \ \ \
\end{equation}
As you can see, $\delta(l)$ is the standardized residual.

To get the distribution of $\delta(l)$ and the value of $p$, we need to use the bootstrap method to generate new samples. That is, generate $M$ random numbers with uniform distribution within $\{1,2,\ldots,M\}$, and construct a new sample with the residual sequence $l_t$ corresponding to the random number, repeat N times (N=1000 in the following), N new bootstrap samples can be obtained.

For each bootstrap sample i ($i=1,2,\ldots,M$), use equation (19) to calculate $\delta_i(l)$, denoted as $\{\delta_1(l),\delta_2 (l),\ldots,\delta_N(l)\}$, without loss of generality, the $\delta(l)$ of the initial sample can be recorded as $\delta_0(l)$. Using the bootstrap method, the empirical distribution of $\delta(l)$ can be obtained.

Since $e_t$ is often a right-skewed distribution, the selected hypothesis $H_1$ of the bootstrap test is $E(e_t)>0$, and the null hypothesis is $H_0 is E(e_t)=0$. Calculate the sample proportion of $\{\delta_1(l),\delta_2(l),\ldots,\delta_N(l)\}$ greater than $\delta_0(l)$, which is the P value of the test. The larger the value of P, the more unable to reject the null hypothesis, that is, the model is considered to have better ES estimation accuracy.

\subsection{Loss function}
When using VaR and ES for risk management, not only need to be concerned about the accuracy of VaR and ES for forecasting, but also need to be concerned about the shape of the entire left tail region of the return distribution. There must be a more accurate understanding of all losses. This article intends to use two risk loss functions to evaluate the prediction of the model.
\subsubsection{Quantile Loss Function (QL)}
Considering the loss function of VaR, this article chooses to use the quantile loss function ((\ref{ql}) formula), which gives a higher weight than VaR. \cite{bib24}
\begin{equation}
\label{ql}
{\rm QL}_t^\alpha=(\alpha-I_t^\alpha)(y_t-{\rm VaR}_t^\alpha)
\end{equation}
\subsubsection{Joint Loss Function (FZL)}
Since there is no separate evaluation of ES loss function , referring to the literature, this paper uses the FZL joint loss function to evaluate the prediction accuracy of ES and VaR for extreme risks.\cite{bib25,bib26}
\begin{equation}
\mathrm{FZL}_{t}^{\alpha}=\frac{1}{\alpha \mathrm{ES}_{t}^{\alpha}} I_{t}^{\alpha}\left( y_{t}-\mathrm{VaR}_{t}^{\alpha}\right)+\frac{\mathrm{VaR}_{t}^{\alpha}}{\mathrm{ES}_{t }^{\alpha}}+\log \left(-\mathrm{ES}_{t}^{\alpha}\right)-1
\end{equation}

\section{Empirical Research}
The indexes selected in this article are Shanghai Stock Index (000001.ss), Shenzhen Index (399001.sz) and Small and Medium-sized Board Index (399005.sz). The sampling time is 3000 days from July 6, 2007 to November 1, 2019. Data source: wind database. Use R language for data processing.
\subsection{Descriptive Statistics}
First, the daily logarithmic rate of return $r_t=lnp_t-lnp_{t-1}$ is calculated, and the basic statistics such as the mean and variance of the corresponding sample rate of return are calculated(table \ref{table1}\footnote{JB-p: P value of Jarque-Bera test, used to test whether a set of samples can be considered as coming from a normal population}. From Table 1, we can see that the mean values of the three samples are relatively close to 0; the variances are roughly the same, and the variance of the small and medium board index is the largest; from the perspective of skewness kurtosis and the JB test p-value, all three samples have peak fat Typical facts with tails and right skewed, and none of them obey normal distribution. From the kurtosis point of view, the excess kurtosis of the three samples are all greater than 0, and the excess kurtosis of the Shanghai Composite Index even reached 4.5072, which implies that we need to use a distribution that can better describe the peak and thick tail to describe the yield.

\begin{table}[H]
	\centering
	\caption{Basic statistics}
	\begin{tabular}{cccccc}
		\toprule
		& mean & variance& skewness &  excess kurtosis & JB-p \\ \midrule
		000001.ss & -0.0001 & 0.0003 & -0.5256 & 4.5072 & 0.0000 \\ 
		399001.sz & -0.0001 & 0.0003 & -0.4660 & 2.9263 & 0.0000 \\ 
		399005.sz & 0.0001 & 0.0004 & -0.5475 & 2.6491 & 0.0000 \\ \bottomrule
	\end{tabular}
\label{table1}
\end{table}

\begin{figure}
	\centering
	\includegraphics[width=0.7\linewidth]{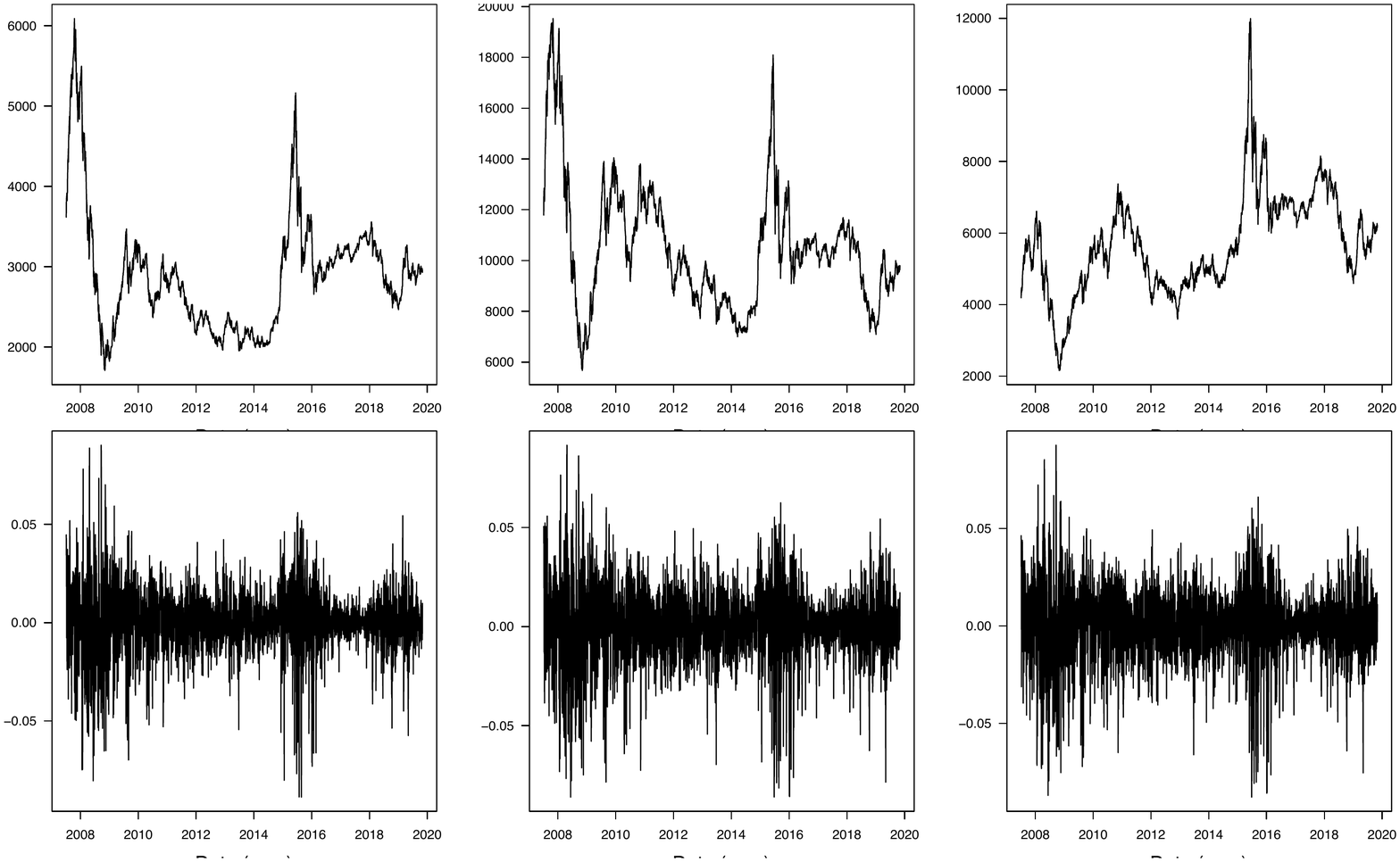}
	\caption{Index trend and return}
\label{fig: index}
\end{figure}

Figure \ref{fig: index} \footnote{From left to right: Shanghai Stock Exchange Index Trend and Return, Shenzhen Index Trend and Return, Small and Medium-sized Board Index Trend and Return}draws a time series diagram of index trend and rate of return. It can be clearly seen that the return rate sequence has volatility clustering.

\subsection{In-sample GAS-ALD fitting}

\subsubsection{Model Fitting}
Fit the GAS-ALD model to the data in the sample period. The results are as follows. It can be seen that among the parameters of the three indices, $a_i $ and $ b_i$ ($i=1,2,3$) are significantly different from 0, indicating that the parameters are time-varying.(table \ref{tablep}\footnote{	K-S test line reports as p value, ***: significant at 99\% level})

In order to test the fitting status of the GAS-ALD model to the distribution of returns, this paper uses the GAS-ALD model for probability integral transformation of the returns, and then performs the K-S test. The data after the probability integral transformation of the original data accepts the null hypothesis of the same distribution as the uniform distribution on (0,1). This shows that the GAS-ALD model can better fit the distribution of returns.

\begin{table}
	\centering
	\caption{Parameter estimation resul}
	\begin{tabular}{ccccc}
		\toprule
		& 000001.ss & 399001.sz & 399005.sz \\ \midrule
		$	\mathbf{\kappa}_1$ & 0 & 0 & 0 \\ 
		$	\mathbf{\kappa}_2$  &$ -0.219^{***}$& $-0.277^{***}$& $-0.277^{***}$\\ 
		$	\mathbf{\kappa}_3$  & 0 & 0 & 0 \\ 
		$a_1$ & $-16.118^{***}$ & $-16.118^{***}$ & $-16.118^{***} $\\ 
		$a_2 $& $-3.945^{***} $& $-3.945^{***}$ & $-3.945^{***}$\\ 
		$a_3$ & $-11.513^{***} $&$ -11.513^{***}$&$ -11.513^{***}$ \\ 
		$b_1$ & $3.897^{***}$ & $3.897^{***}$ & $3.897^{***}$ \\ 
		$b_2$ & $2.883^{***}$& $2.593^{***}$ & $2.593^{***}$ \\ 
		$b_3$ & $3.897^{***}$ &$ 3.897^{***}$ &$ 3.897^{***}$ \\ 
		K-S test & 0.737 & 0.751 & 0.426 \\ \bottomrule
	\end{tabular}
\label{tablep}
\end{table}

\subsubsection{Analysis of dynamic changes of parameters and moments}
\begin{figure}[H]
	\centering
	\begin{minipage}[t]{0.48\textwidth}
		\centering
		\includegraphics[width=0.7\linewidth]{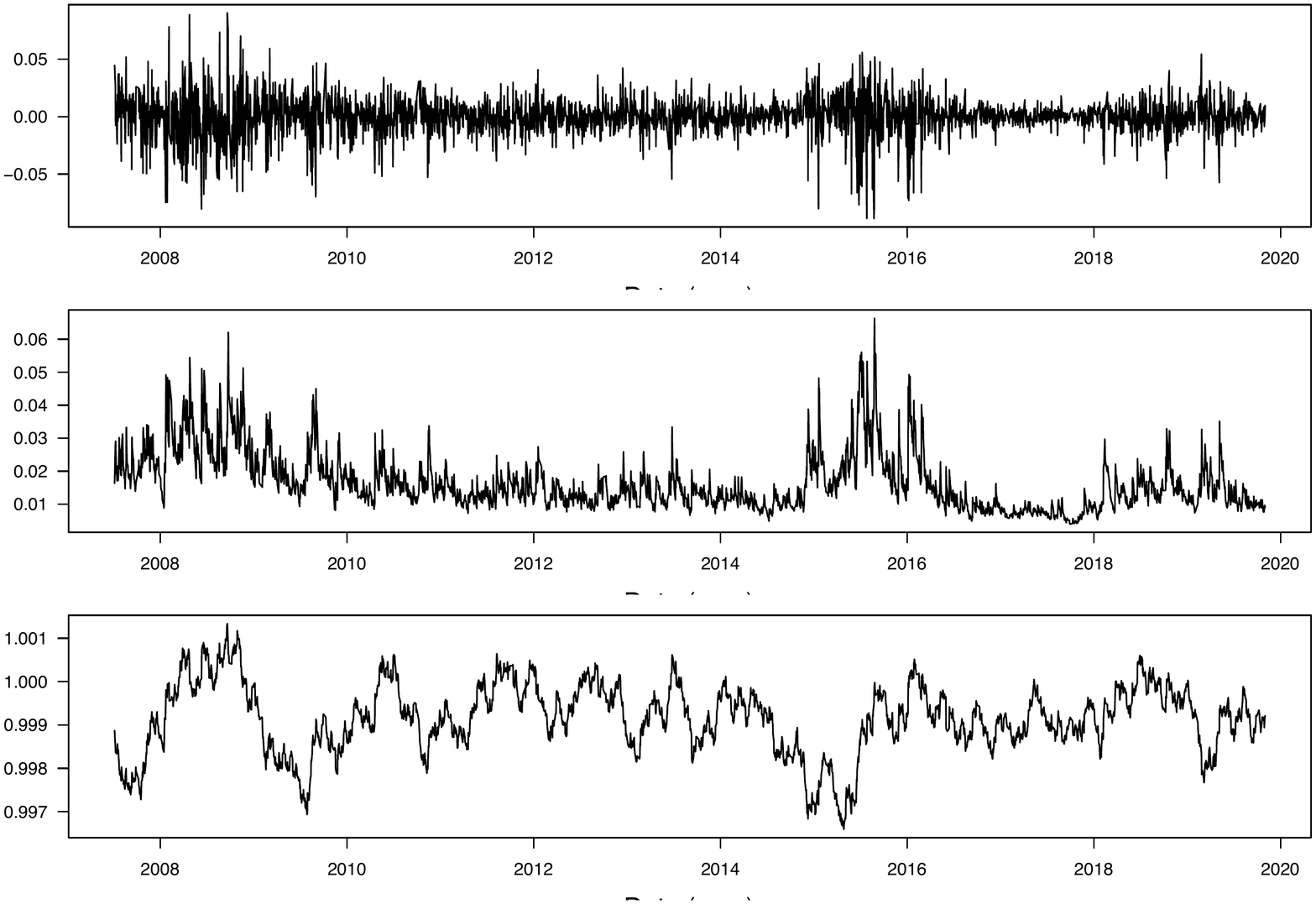}
		\caption{Shanghai Composite Index Dynamic Parameters}
		\label{Shanghai Stock Index Dynamic Parameters}
	\end{minipage}
	\begin{minipage}[t]{0.48\textwidth}
		\centering
		\includegraphics[width=0.7\linewidth]{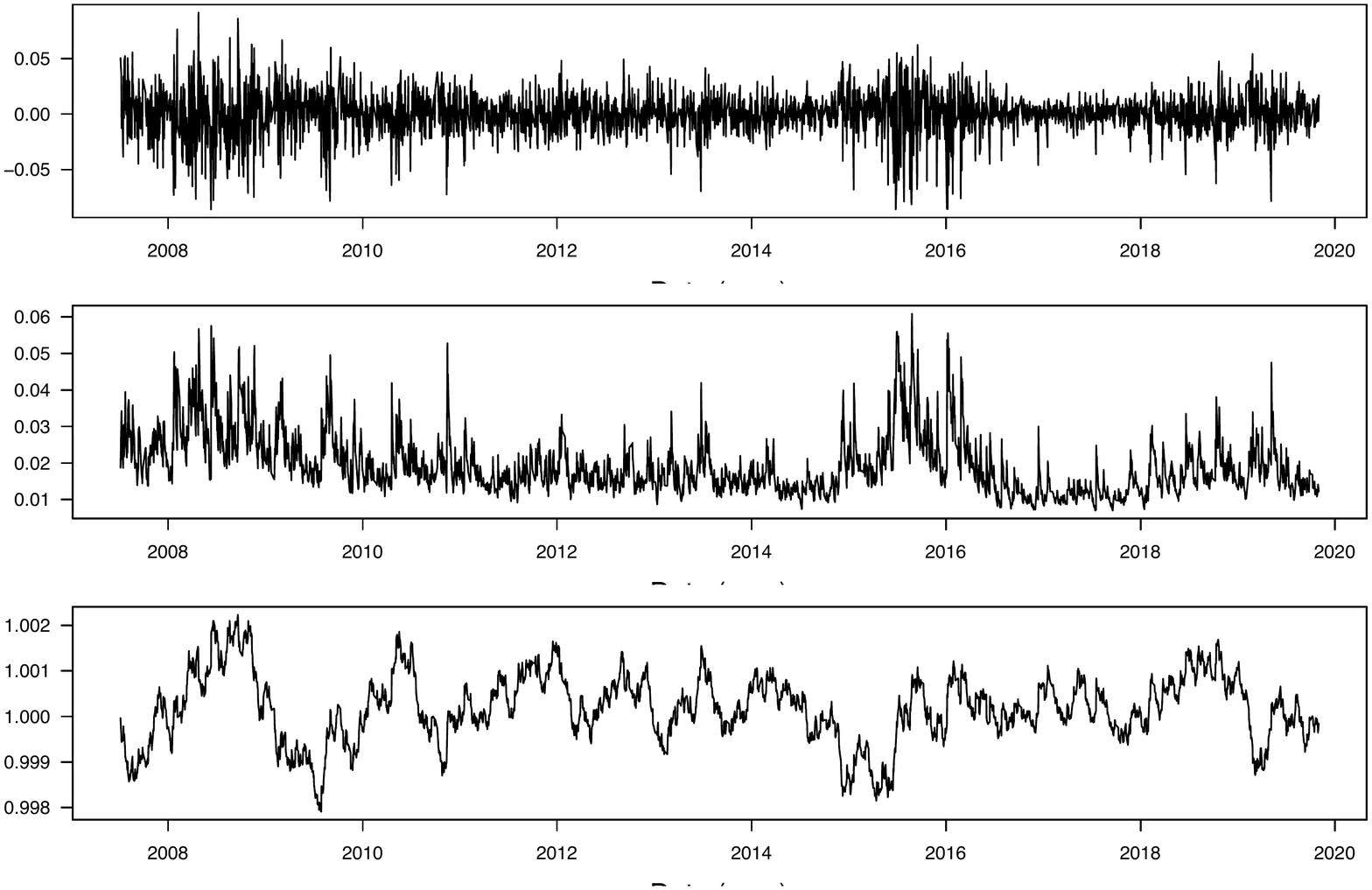}
		\caption{Shenzhen Index Dynamic Parameters}
		\label{Shenzhen Index Dynamic Parameters}
	\end{minipage}
	\begin{minipage}[t]{0.48\textwidth}
		\centering
		\includegraphics[width=0.7\linewidth]{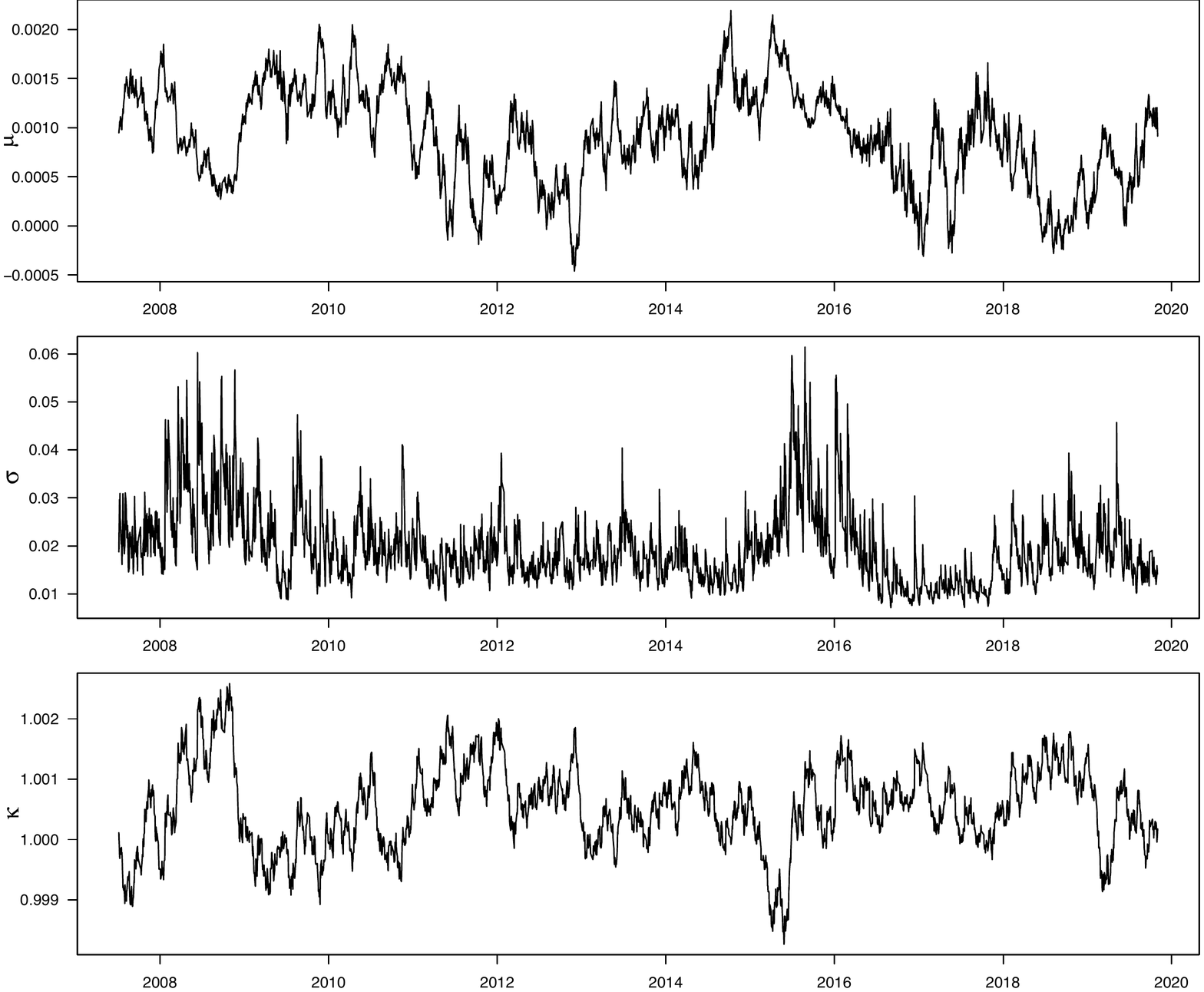}
		\caption{ Small and medium board index Dynamic Parameters}
		\label{Small and Medium Board Index Dynamic Parameters}
	\end{minipage}

\end{figure}

This article draws the time-varying trajectories of three parameters (Figure \ref{Shanghai Stock Index Dynamic Parameters}, Figure \ref{Shenzhen Index Dynamic Parameters} and Figure \ref{Small and Medium Board Index Dynamic Parameters})\footnote{The three pictures from top to bottom are: $\mu_{t}$,$\sigma_{t}$,$p_{t}$.} It can be seen that the parameter trends of the three index samples are basically the same. $\sigma_t$ has the strongest aggregation, and there are two obvious peaks. And noticed that after the peak of $\sigma_t$ appears, it is often accompanied by a rapid decline in $p_t$ and $\mu_t$.

\begin{figure}[H]
	\centering
	\begin{minipage}[t]{0.48\textwidth}
		\centering
		\includegraphics[width=0.7\linewidth]{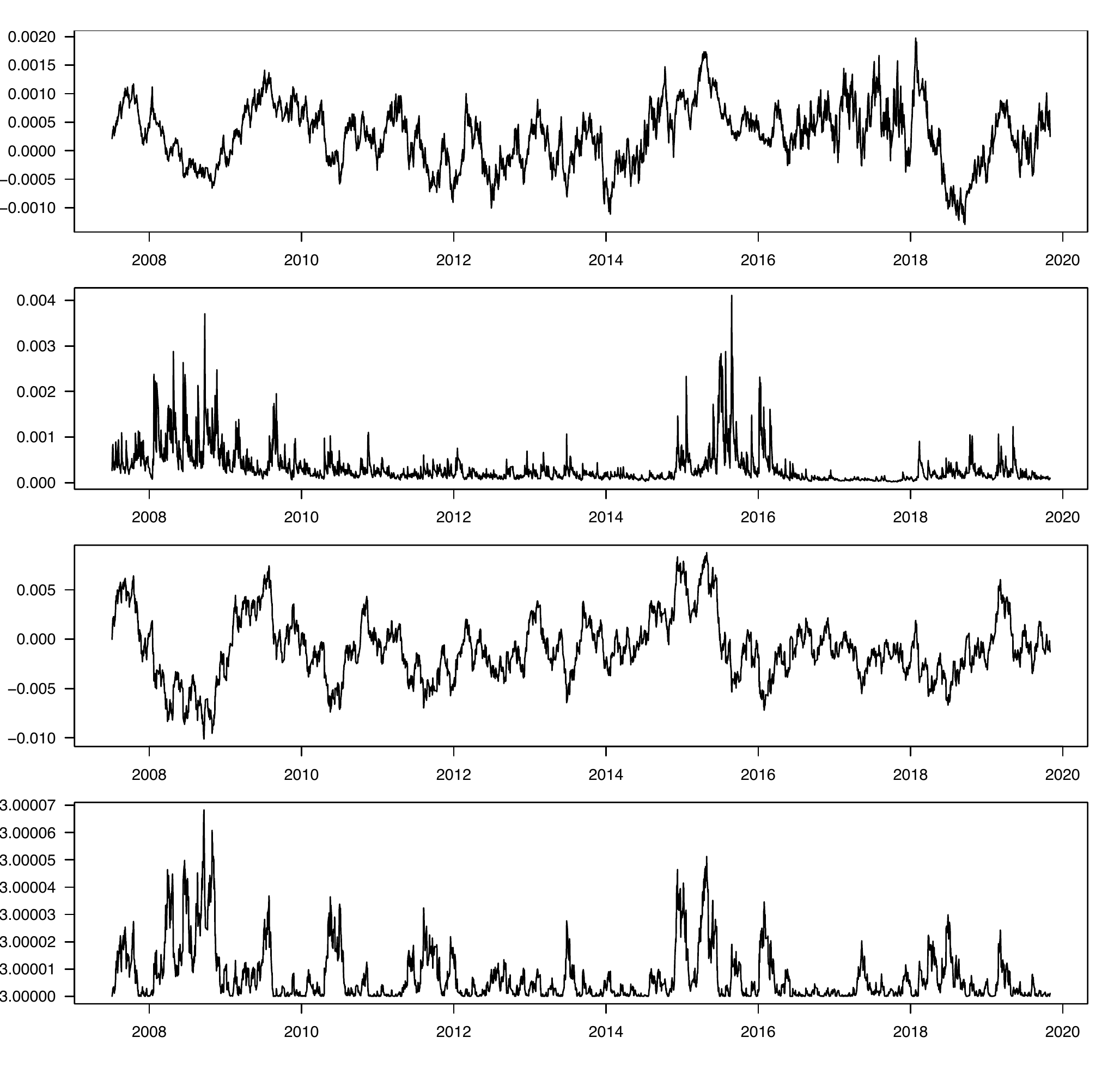}
		\caption[上证指数矩结果]{Shanghai Composite Index Moment Results}
		\label{Shanghai Stock Index Moment Results}
	\end{minipage}
	\begin{minipage}[t]{0.48\textwidth}
		\centering
		\includegraphics[width=0.7\linewidth]{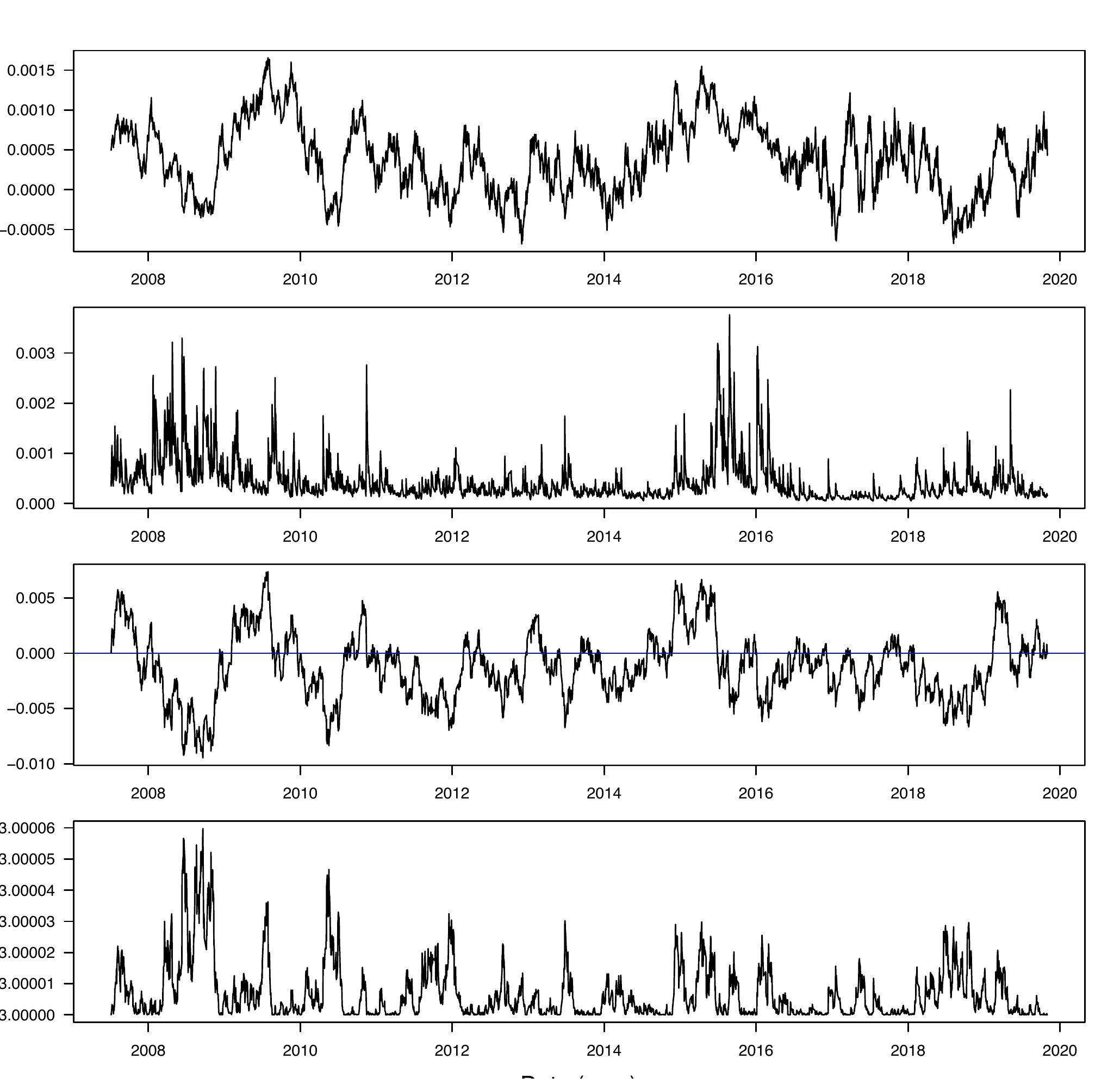}
		\caption[深圳指数矩结果]{Shenzhen Index Moment Results}
		\label{Shenzhen Index Moment Results}
	\end{minipage}
	\begin{minipage}[t]{0.48\textwidth}
		\centering
		\includegraphics[width=0.7\linewidth]{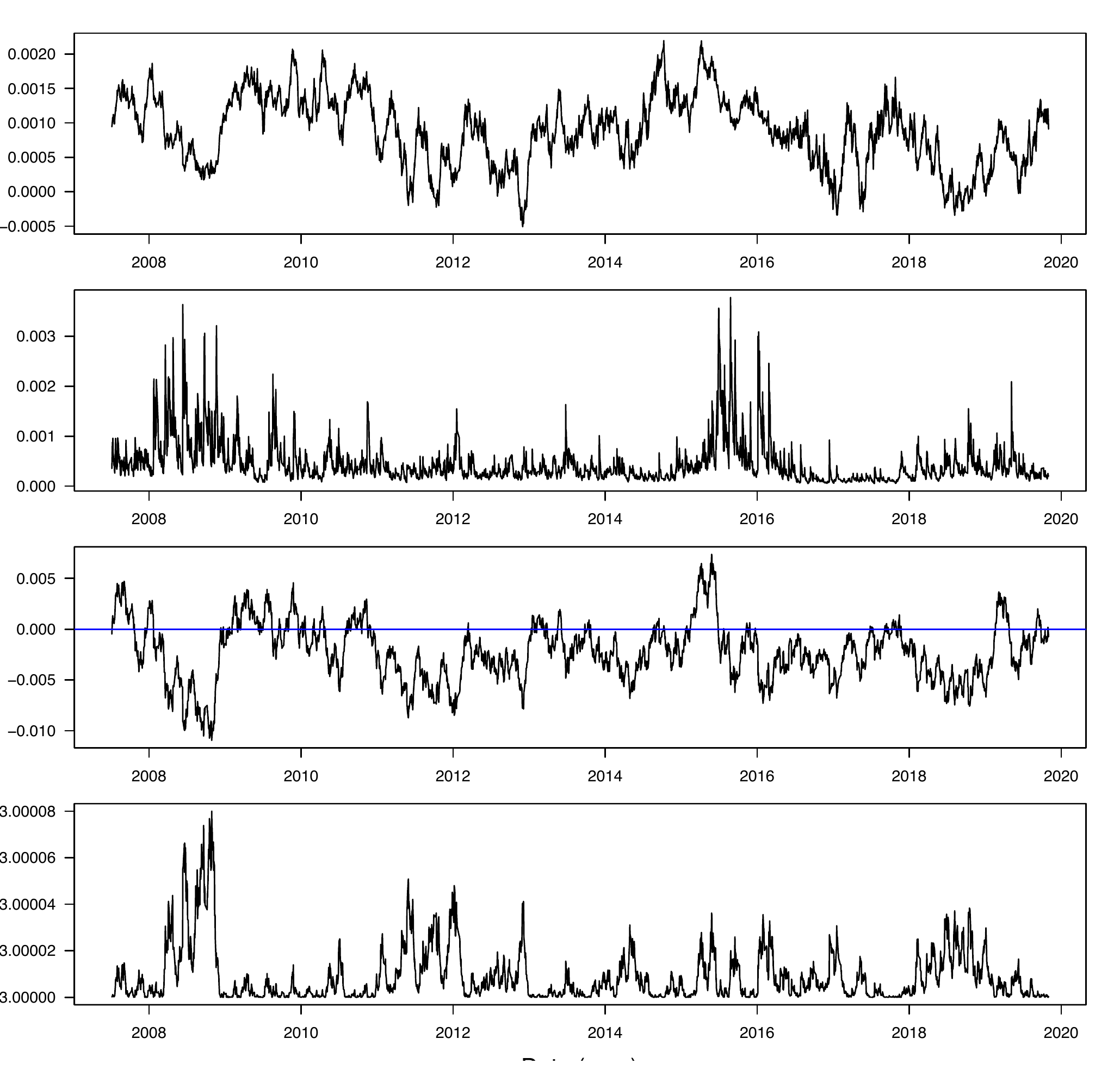}
		\caption[中小板指数矩结果]{Small and medium board Index Moment Results}
		\label{Small and Medium Plate Index Moment Results}
	\end{minipage}

\end{figure}

Calculate conditional mean, conditional volatility, conditional skewness and conditional kurtosis based on dynamic parameters. The results are shown in Figure \ref{Shanghai Stock Index Moment Results}, Figure \ref{Shenzhen Index Moment Results} and Figure \ref{Small and Medium Plate Index Moment Results}\footnote{The four graphs from top to bottom are: conditional mean, conditional volatility, conditional skewness and conditional kurtosis} Conditional volatility and conditional kurtosis have the strongest aggregation effect. The second is the mean and kurtosis. It shows that each moment has obvious time-varying characteristics and aggregation characteristics. From the perspective of linkage, every sudden increase in volatility will result in a sharp drop in skewness and an increase in kurtosis. This shows that when the volatility rises, there will be a more obvious sharp peak and thick tail right deviation, which will lead to increased tail loss. And it can be seen that the change trends of the moments of the three indices are roughly the same, which shows that the correlation of the moments of the three indices is relatively large.

\subsubsection{Time point analysis}
\begin{figure}[H]
	\centering
	\begin{minipage}[t]{0.48\textwidth}
		\centering
		\includegraphics[width=0.6\linewidth]{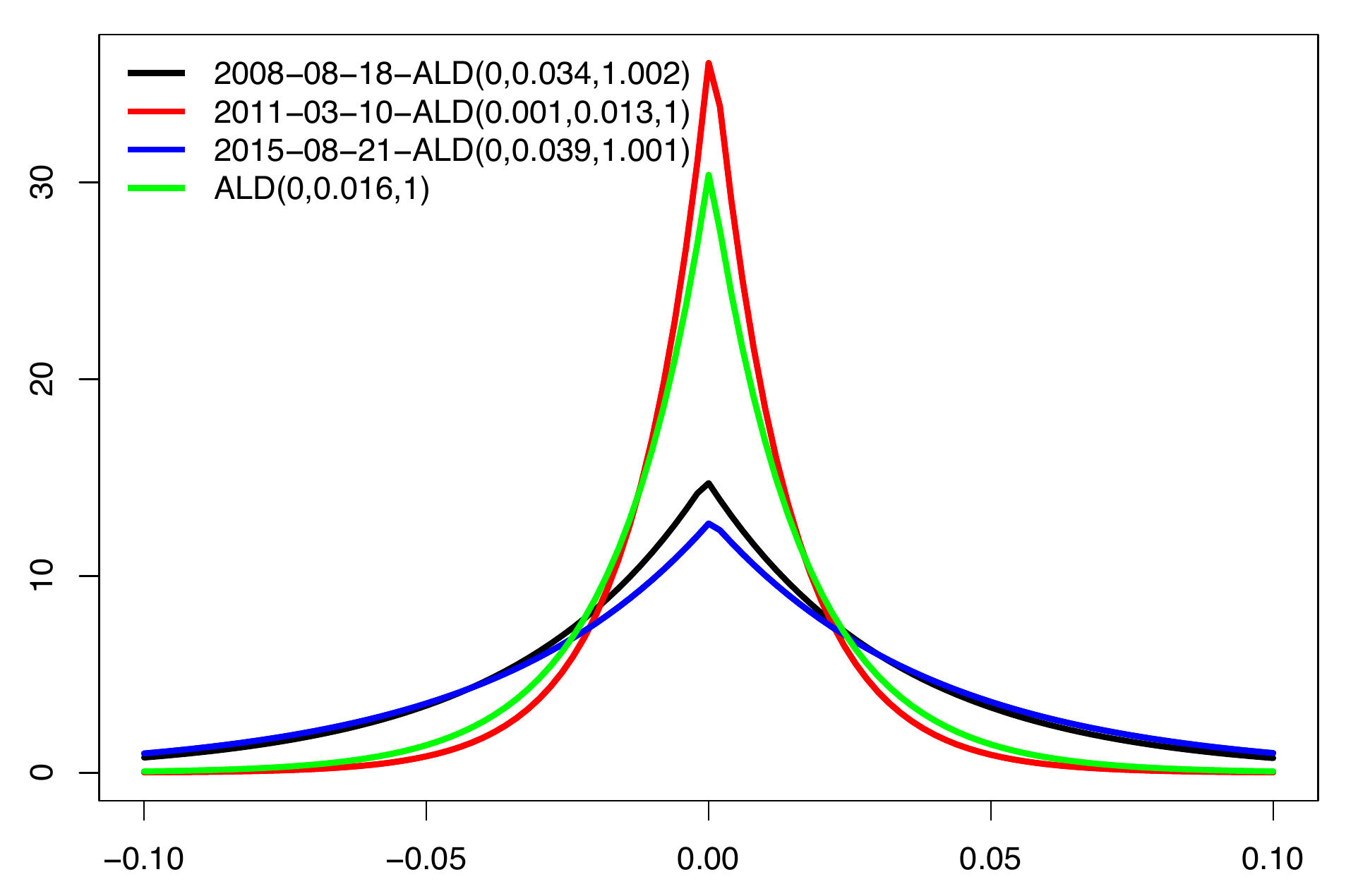}
		\caption[上证参数]{Time distribution of Shanghai stock index return rate}
		\label{上证指数收益率时点分布}
	\end{minipage}
	\begin{minipage}[t]{0.48\textwidth}
		\centering
		\includegraphics[width=0.6\linewidth]{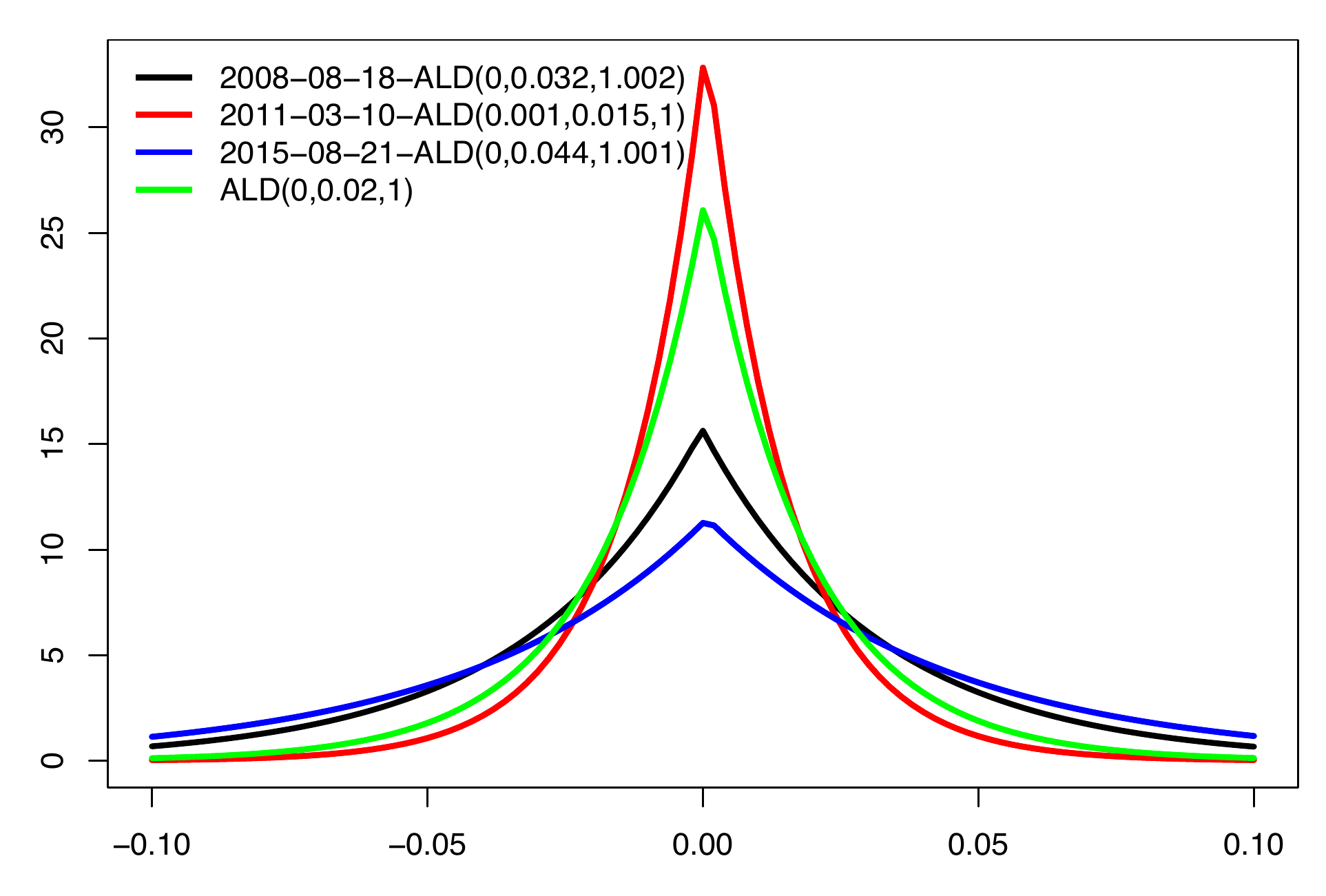}
		\caption[深圳参数]{Time distribution of Shenzhen index return}
		\label{深圳指数收益率时点分布}
	\end{minipage}
	\begin{minipage}[t]{0.48\textwidth}
		\centering
		\includegraphics[width=0.6\linewidth]{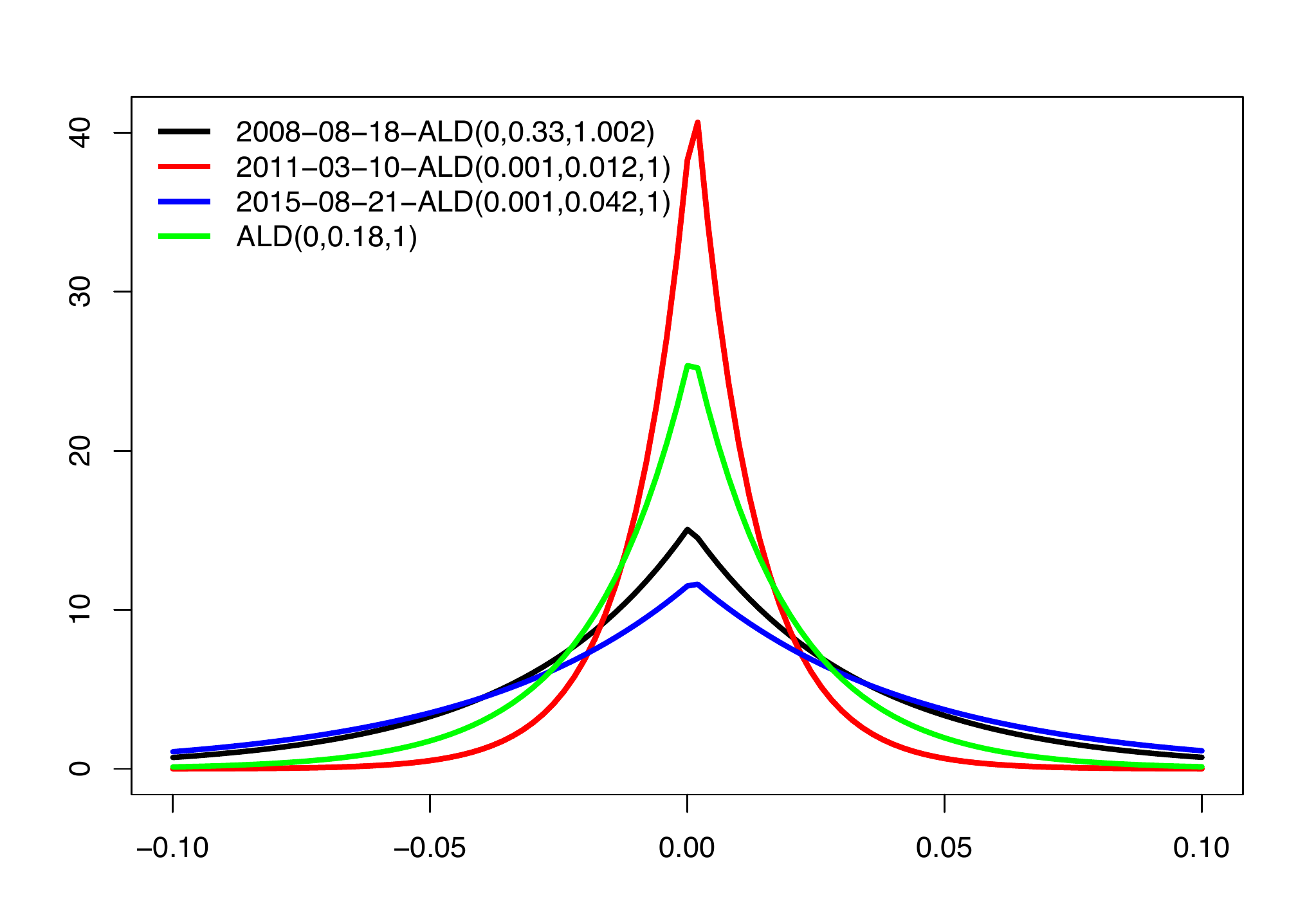}
		\caption[中小板指数收益率时点分布]{Time distribution of small and medium board index returns}
		\label{中小板指数收益率时点分布}
	\end{minipage}

\end{figure}

Consider three time points for comparative analysis. The time points selected in this article are 2008-08-18, 2011-03-10 and 2015-08-21, which represent the period of the subprime mortgage crisis, the rising period, and the period of leverage stock disasters, respectively. Plot the return distribution of each index at each time point. As shown \ref{上证指数收益率时点分布}, \ref{深圳指数收益率时点分布} and \ref{中小板指数收益率时点分布}\footnote{The green line is the asymmetric Laplace distribution density obtained by the full sample fitting, and the black line, red line and blue line are the asymmetric Laplace distribution density at a specific time point}.
\begin{figure}[H]
	\centering
	\includegraphics[width=0.5\linewidth]{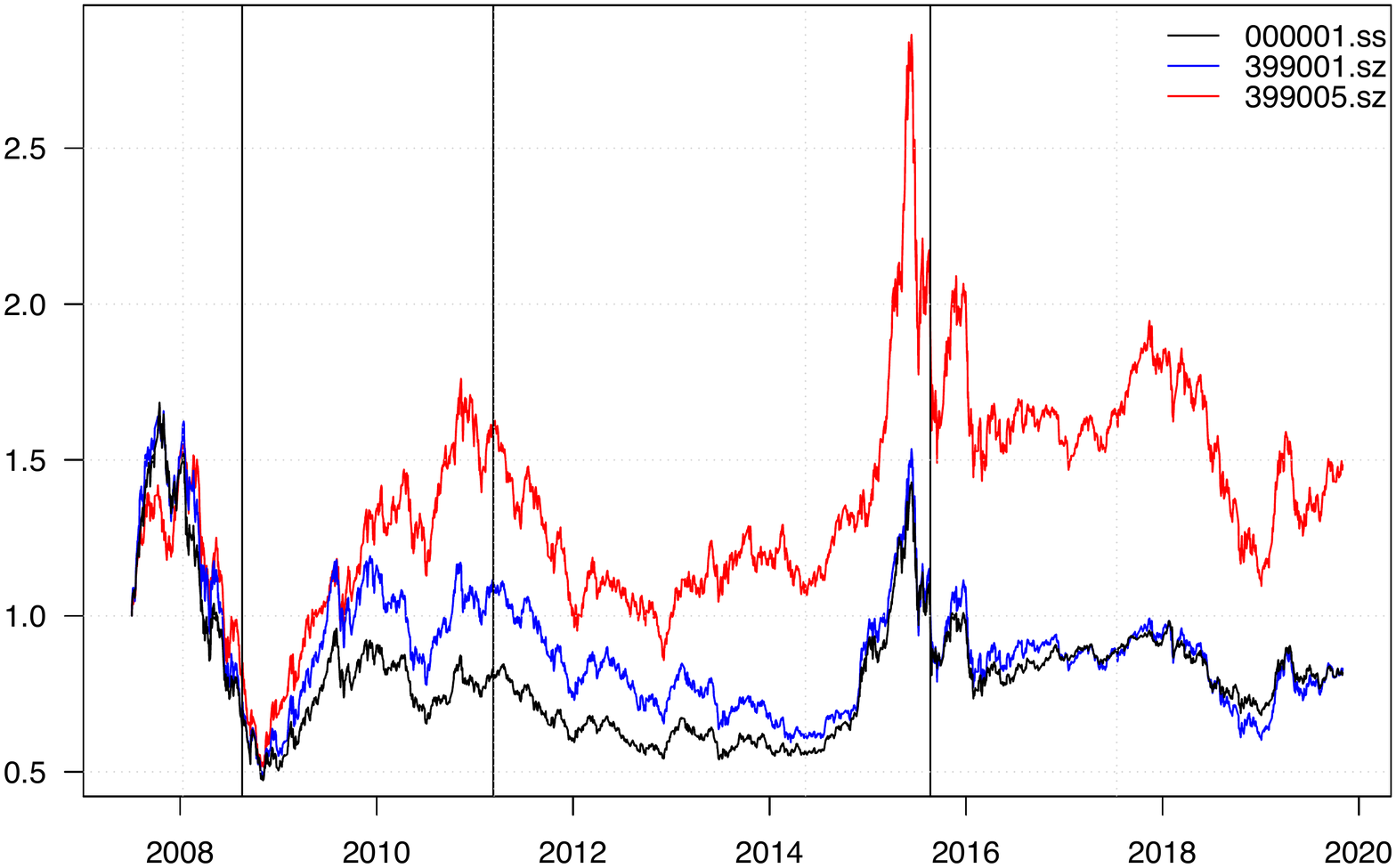}
	\caption{Time selection}
	\label{fig:}

\end{figure}

  It can be found first. During the rising period, the distribution of the return rate is sharp and thin compared to the full sample period, indicating that most of the return is concentrated near 0, and the value of VaR and ES can be appropriately lowered at this time. During the subprime mortgage crisis and leveraged stock crisis, compared with the full sample, the tail thickness increased, indicating that the probability of loss at this time is much greater than that of the full sample and the rising period. At this time, the risk exposure faced by investors will greatly increase. In addition, when a leveraged stock disaster occurs, the thickness of the tails of the three index yield distributions is greater than that of the subprime mortgage crisis, indicating that the risk exposure faced by investors during the leveraged stock disaster is greater than the risk exposure faced by the subprime mortgage crisis. This is because in 2015, a large number of investors increased their investment leverage through borrowing, resulting in losses that they faced when the stock market plummeted far beyond their own capabilities.

\subsection{Rolling forecast VaR and ES}
This article uses three types of models as a comparison. The first type is historical simulation method and fixed parameter ALD distribution. Among them, the historical simulation method uses short-term (25-day) sample quantiles, and the second type is other distributions under the GAS framework. Normal distribution (GAS-norm), t distribution (GAS-t) and partial t distribution (GAS-skst). The third type is GARCH model. This article uses APARCH model with partial t distribution error term (APARCH-skst) As a control group.

We selects the first 1500 days of the sample as the training group, rolls forward forecast 1500 days, re-estimates the model every 5 days, and calculates the VaR and ES of 1\%.
\subsubsection{Backtesting}
Table \ref{var} reports the results of the VaR backtest. Combining the UC, CC, and DQ test results, GAS-norm and GAS-t performed the worst. This may be due to the fact that the normal distribution and t distribution cannot describe the asymmetry and biased characteristics of their returns. This shows that when calculating VaR, the model needs to be able to describe the asymmetry and biased characteristics of the rate of return. The GAS-ALD model has the best performance in the UC, CC and DQ tests of the Shanghai Stock Index and the Small and Medium-sized Board Index; the GAS-ALD model has the best performance in the UC and CC tests of the Shenzhen Index, and the DQ test is slightly worse. The GAS-skst model, which shows that the Shenzhen Index is more suitable to use the GAS-skst model to calculate the VaR of 1\%.

Table \ref{est} reports the ES backtest results. It was found that GAS-norm, GAS-t and GAS-skst performed poorly in calculating ES, and all rejected the null hypothesis. The ES backtest of the Shanghai Composite Index shows that the GAS-ALD model is slightly worse than APARCH-skst; the Shenzhen Index and the small and medium-sized board index have the best performance in the ES backtest of the GAS-ALD model.

\begin{table}[H]
	\centering
	
	\caption{VaR Backtesting}
	\resizebox{\textwidth}{15mm}{
		\begin{tabular}{c|ccc|ccc|ccc}
			\toprule
			& \multicolumn{3}{|c|}{ 000001.ss }& \multicolumn{3}{|c|}{399001.sz }& \multicolumn{3}{|c}{399005.sz} \\ \midrule
			& UC.pval & CC.pval & DQ.pval & UC.pval & CC.pval & DQ.pval & UC.pval & CC.pval & DQ.pval \\ 
			Historical Simulation & 0.131 & 0.334 & 0.000 & 0.405 & 0.334 & 0.000 & 0.611    & 0.041 & 0.000 \\ 
			Fixed Parameter ALD  & 0.217 & 0.252 & 0.000 & 0.142 & 0.035 & 0.000 & 0.702 & 0.046 & 0.000 \\ 
			GAS-norm & 0.003 & 0.006 & 0.001 & 0.01 & 0.022 & 0.004 & 0.001 & 0.002 & 0.006 \\ 
			GAS-t & 0 & 0.001 & 0.004 & 0 & 0 & 0 & 0 & 0 & 0 \\ 
			GAS-skst & 0.003 & 0.006 & 0.017 & 0.054 & 0.11 & 0.385 & 0.001 & 0.003 & 0.004 \\ 
			GAS-ALD & 0.142 & 0.252 & 0.534 & 0.142 & 0.252 & 0.241 & 0.089 & 0.17 & 0.238 \\ 
			APARCH-skst & 0.032 & 0.067 & 0 & 0.089 & 0.17 & 0.006 & 0.089 & 0.17 & 0.007 \\ \bottomrule
		\end{tabular}
	}
	\label{var}
\end{table}
\begin{table}[H]
	\centering
	\caption{ES Backtesting}
	\begin{tabular}{ccccc}
		\toprule
		ES-1\% & 000001.ss & 399001.sz & 399005.sz \\ \midrule
		Historical Simulation & 0.014 & 0.006 & 0.001 \\ 
		Fixed Parameter ALD & 0 & 0.247 & 0.032 \\
		GAS-norm & 0 & 0 & 0 \\ 
		GAS-t & 0 & 0 & 0 \\
		GAS-skst & 0 & 0 & 0 \\ 
		GAS-ALD & 0.271 & 0.527 & 0.533 \\ 
		APARCH-skst & 0.658 & 0.373 & 0.339 \\ \bottomrule
	\end{tabular}
	\label{est}
\end{table}
\subsubsection{Loss function}
Use the predicted VaR and ES results to calculate the QL loss function and FZL loss function. The results are shown in Table \ref{table5}\footnote{The QL loss function results are all multiplied by $10^3$, and the reported value is the average}.

For the QL loss function, the GAS-ALD model has the smallest average QL in the Shanghai Stock Index and the Shenzhen Index, indicating that among the compared models, GAS-ALD has better prediction accuracy for VaR prediction, while in the small and medium board In the index, the average QL of APARCH-skst is the smallest, followed by the average QL of GAS-ALD. This article believes that this is because APARCH-skst cannot dynamically describe the changes in distribution, but it can describe the asymmetric effect of the return rate. There are often obvious asymmetric effects in the small and medium board index, which makes APARCH-skst's prediction accuracy for VaR Slightly higher than the GAS-ALD model.

For the FZL loss function, the GAS-ALD model has the smallest average FZL value, indicating that the GAS-ALD model has better ES prediction accuracy than the control model.

\begin{table}[H]
	\centering
	\caption{Loss Function}
	\begin{tabular}{c|cc|cc|cc}
		\toprule
		& \multicolumn{2}{|c|}{000001.ss} & \multicolumn{2}{|c|}{399001.sz} & \multicolumn{2}{|c}{399005.sz} \\ \midrule
		Loss Function & QL & FZL & QL & FZL & QL & FZL \\ 
       Historical Simulation & 7.13 & -2.629 & 7.428 & -2.594 & 7.5 & -2.611 \\ 
		Fixed Parameter ALD& 7.198 & -2.627 & 7.501 & -2.597 & 7.399 & -2.589 \\ 
		GAS-norm & 5.599 & -2.921 & 6.693 & -2.621 & 6.812 & -2.582 \\ 
		GAS-t & 5.852 & -2.768 & 7.081 & -2.491 & 7.29 & -2.392 \\ 
		GAS-skst & 6.301 & -2.711 & 7.74 & -2.444 & 7.749 & -2.436 \\ 
		GAS-ALD & 5.543 & -2.994 & 6.631 & -2.736 & 6.716 & -2.706 \\ 
		APARCH-skst & 5.453 & -2.984 & 6.839 & -2.833 & 6.167 & -2.614 \\ \bottomrule
	\end{tabular}

	\label{table5}
\end{table}
\section{Conclusion}

In this paper, by improving the asymmetric Laplace distribution so that its parameters are all time-varying, the GAS-ALD model is proposed and used to conduct empirical research on China’s Shanghai Stock Exchange Index, Shenzhen Index and Small and Medium-sized Board Index. The following conclusions:

1) The distribution of logarithmic returns of Shanghai Composite Index, Shenzhen Index and Small and Medium-sized Board Index conforms to the GAS-ALD model, and its conditional mean, conditional volatility, conditional skewness and conditional kurtosis are all time-varying and clustering. If the normal distribution is used Waiting for the static distribution will lead to failure to capture the time-varying characteristics of the moment.

2) UC test, CC test, DQ test and bootstrap test are used to test the risk prediction ability of GAS-ALD model. Compared with GAS-norm and GAS-t, which are equal to the distribution under the GAS framework and APARCH-skst, the GAS-ALD model has a better feasibility on VaR and ES of 1\%. And the use of QL loss function and FZL loss function shows that the GAS-ALD model has better prediction accuracy.

In summary, GAS-ALD can better describe the distribution of peaks and thick tails, biased, and left-to-right asymmetry, and has good accuracy in the prediction of VaR and ES. These conclusions can provide some help in risk measurement for participants in the capital market.

As a means of describing the distribution, GAS-ALD can be used as the marginal distribution of returns to fit Copula, and to measure the VaR and ES of the portfolio. In the future we will involve more research in this area.

\end{document}